\documentstyle[aps,prl,epsf,psfig,twocolumn,floats]{revtex}
\begin{document} 
\draft
\twocolumn[\hsize\textwidth\columnwidth\hsize\csname @twocolumnfalse\endcsname
\title{Kinetics of condensation of flexible polyelectrolytes in poor solvents:
effects  of solvent quality, valence and size  of counterions }
\author{Namkyung Lee and  D. Thirumalai}
\address{Institute for Physical Science and Technology, 
University of Maryland, College Park, Maryland  20742} 
\date{\today} 
\maketitle 
 
\begin{abstract}
 The collapse kinetics of strongly charged polyelectrolytes in
poor solvents is  investigated by Langevin simulations and scaling arguments.
The rate of collapse  increases  sharply as the valence of
counterions, $z$, increases from one to four. The \textit{combined system}
of the
collapsed  chain and the condensed counterions forms
a Wigner crystal
when the solvent quality is not too poor provided  $z \geq 2$. For very poor
solvents the morphology of the collapsed structure resembles
a Wigner glass. For a fixed $z$ and quality of the solvent the 
efficiency of collapse decreases dramatically  as the size of the counterion
increases.
A valence dependent diagram of states in poor solvents  is derived.
\end{abstract}
]
\narrowtext
Upon condensation of DNA into toroidal structures
its   volume decreases by nearly 
four orders of magnitude \cite{bloomfield1}.
Counterion condensation,  which leads to attractive interactions
between the segments of the polyanion chain, is a key factor
in determining DNA collapse \cite{ha}.
Experiments show that counterions with trivalent (or greater) valence, $z$, are often required to 
induce chain collapse\cite{bloomfield1}.   
Precipitation (requiring chain contraction)  in aqueous solution of a highly 
charged flexible polyelectrolyte (PE)  (polystyrene sulfonate)
occurs only  when counterions with $z=3$ or $4$ are added\cite{olvera}.
The collapse of DNA (a stiff chain)  depends on  the interplay of  
several effects  such as bending rigidity, valence and size of 
counterions, and hydration forces\cite{bloomfield1}. Since  intrinsic stiffness is not relevant for 
flexible PE  the collapse transition may be  simpler to describe  than in 
stiff PE.  However, because of competing effects of the bare electrostatic interactions,  counterion condensation 
and  the solvent quality  the phase diagram of highly charged flexible PEs is quite complicated\cite{schiessel}. 
Simulations\cite{kremer,winkler}, scaling arguments\cite{schiessel,de-gennes,barrat},  and theory\cite{bril} are beginning to provide a picture of the possible structures that emerge in highly charged polyelectrolytes  in which correlations induced by counterions play  a crucial role\cite{schiessel,kremer,winkler,bril,khokhlov}.

In this letter we describe   the dynamics of collapse of strongly charged 
flexible PE
in poor solvents with  emphasis on the mechanisms of collapse  and 
their dependence on the valence, size  of counterions and solvent quality. 
We consider  polyelectrolyte chains that are  highly charged so that $f(l_B/b)^2 >1$ where f(=1 in the examples considered here) is the fraction of charged monomers, $l_B=e^2/4\pi \epsilon k_{B} T$ is the Bjerrum length, with T being temperature, $\epsilon$ is the dielectric constant of the solvent, and $b$ is the size of the monomers. 
The major results of this study, which are obtained using a combination of Langevin simulations and scaling arguments, are:

(a)  The mechanisms and     kinetics  of collapse   dramatically 
depend on the valence $z$  of the counterions. For all range of 
parameters the slowest kinetics is found for $z=1$, and generally the 
rate  of globule  formation increases with $z$. 

(b) The equilibrium structure of the collapsed   polyelectrolyte  depends critically  on the solvent quality. If the solvent is not too poor then the conformation 
of the counterion and the charged polymer is a Wigner crystal. For $z\geq 2$,
 we find a BCC-like crystal. When the solvent quality is very poor  amorphous
 structures  (Wigner glasses) are  formed. 

(c) For a fixed   solvent quality, the efficiency of chain collapse depends dramatically on the size of 
the counterions. The rate of approach to the globular conformation increases
as the size of the counterion decreases at a fixed value of $z$.

The polyelectrolyte is taken to be a  flexible chain consisting of $N$
 monomers
each with  a charge of $-1$ (measured in units of $e$).   Successive monomers are connected by 
an elastic spring with spring constant $3 k_BT/b^2$.  
The valence  of the cations  varies from $z=1-4$ and their 
density  is determined by the condition of charge neutrality. 
The non-electrostatic interaction between the  particles
 (monomer or counterion) 
i and j   with radii $\sigma_i$ and $\sigma_j$  respectively 
that are   separated by $r_{ij}$ is taken to be 
$ H_{LJ} (r_{ij})/k_BT = \epsilon_{LJ}[(\frac{r_o}{r_{ij}})^{12} - 2(\frac{r_o}{r_{ij}})^6] $
where  average bond length  $r_o = \sigma_i+\sigma_j $.
The quality of the solvent is expressed in terms of the second virial 
coefficient $v_2 = \int_v d^3 \bf{r}\it (1-e^{-\frac{H_{LJ}}{k_BT}})$ which is negative for poor solvents.

 The chain and the counterions are placed in a  truncated octahedron (Wigner--Seitz cell of BCC) 
unit cell with the lattice constant $L$ and the  volume of the unit cell is   $V =\frac{1}{2} L^3$.  
The long range nature  of the Coulomb interaction  is treated by accounting for
interactions with the image charges in a  periodically replicated  system\cite{ewald}.
The computations  were done with $N=120$ ($L=40\sigma_m$) or $N=240$ ($L=80 \sigma_m$). 
Most of  the results presented here are for the latter case so that 
the monomer density $\rho_m = N/V = 9.36 \times 10^{-4} \sigma_m^{-3}$, where $\sigma_m (=b/2)$ 
is the radius of the monomer.  We start from  either 
$\Theta$-conditions or from high temperature  and quench the temperature 
below  Manning\cite{manning} condensation condition.  We have verified that 
 the results (collapse mechanisms  and the morphology of the 
collapsed structures) \it do not depend \rm on the initial conditions. 
 The chain  dynamics
 is  described  by the Langevin equation.  The equations of motion are integrated with a step size $= 10^{-3} \tau$ where  the unit of time 
$\tau = b^2/D = b^2 \zeta /k_BT$, where $\zeta$ is the friction coefficient. 
In order to obtain results  with  relatively small errors  averages of various 
quantities are taken over  at least 40 independent initial conditions. 
  
\begin{figure}[h]
\leavevmode\centering\psfig{file=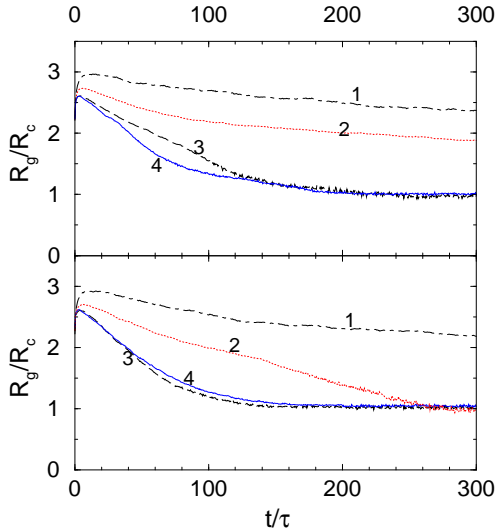,width= 6.5 cm}
\caption{The time dependence of the radius of gyration following a quench from $\Theta$-solvent to poor solvent condition.  The value of $N=240$ and $l_B=5.3b$.  Fig.(1a) is for $v_2= -0.06b^3$ where as Fig.(1b) is for $v_2= -3.69b^3$. The numbers on the curves  denote the valence of the counterions.}
\label{fig1}
\end{figure}

  The time dependence
of $<R_g(t)>/R_c$ for various values of $z$, where $R_c$ is the size of the
compact globule,  shows that for  $v_2=-0.06b^3, l_B = 5.3b$ (Fig.(1a))  
only  counterions  with $z=3$ or $4$ 
induce  globule formation on the time scale of the computations.
The structure of the  chain  with $z=2$ is  extended  at $t\approx 300\tau$.  When the quality of the solvent is 
 poorer ($v_2=-3.69b^3$) then the condensation occurs as long  as $z \geq 2$ (see Fig.(1b)).  Monovalent counterions are inefficient in causing chain collapse. The rate of approach to globular conformation is largest  for $z=4$ and it decreases as  the valence  is reduced. 
 Although the final equilibrium structure is expected to be the same, the kinetics is sensitively dependent on the valence of the counterions.  
For $z\geq 3$, the difference in kinetics  becomes negligible 
as the solvent  quality decreases.  

 The approach to the globular structure   occurs in  roughly three stages for all $z$.
 In the initial stage  counterions  condense onto the  polyanion. 
This takes place on a diffusion controlled time scale
$\tau_{cc} \approx \rho_{m}^{-2/3}/D$ 
which for our simulation is   $\approx 25\tau$. 
We calculated the time required for $90\%$ of counterions to condense.  
The time scales are determined only by $\rho_m$ and are \textit{independent} 
of $z$ and $v_2$.
In the second stage
globular clusters containing both monomers and counterions form.
 The clusters, the number of which is  dependent on $z$ and $v_2$, are connected 
by ``strings''  giving rise to  pearl-necklace structures (see below). This process occurs in the time regime $\tau_{cc} < t < \tau_{CLUST}$. For $t\gg\tau_{CLUST}$,  coarsening of the clusters takes place till the equilibrium  globular structure is formed\cite{LT}.

The  variations in the collapse  kinetics with $z$ may be understood in terms of formation of  clusters (or pearls) and subsequent growth (or merging) of such clusters into a compact globule\cite{homo}.  When  monovalent counterions are condensed they combine with the charges on the backbone of the  polyanion to form ion pairs. 
Even though  the condensed counterions are mobile  along the polyanion backbone, upon  condensation  randomly oriented  dipoles of magnitude $p\approx eb$
are formed.  In order to form  contacts between the segments of the chain the attractive energy between the dipoles must  exceed $k_BT$.
Thus, the  range of  attractive interactions $r_t\approx (l_Bb^2)^{1/3} < l_B$. Since  $r_t/l_B < 1$  we expect cluster formation to occur only on short length scales.  This generates a  large number of clusters and  their subsequent merger (nucleation) to form a fully compact globule is expected to be  slow.
In fact,  we do not see globule formation with $z=1$ on the time scales of our simulation for any value of $v_2$.

A very different mechanism emerges when chain collapse  is induced by multivalent cations.  When  the Manning parameter $\xi_{M} = l_B/b > 1/z$
some of the ions are  condensed\cite{olvera}. 
 A counterion can locally 
neutralize the  negative charge on the monomer,  and the resulting  
bound  species  has an effective charge $(z-1)e$.
Bare monomer charges that are separated by a large distance along the contour can be attracted to the positive charge -  a process referred to as  ``ion-bridging''\cite{olvera}.  The range over which such attractive interactions are effective  clearly 
 increases  with $z$. Since the size (and hence the 
number of clusters ) of the clusters is controlled by the length 
of the contour over which  ionbridges form, it  follows that 
the efficiency of collapse should  increase with $z$. 
The time dependence of the correlation function  
$c(t) = \sum_{i<j, |i-j| \geq 5} \Theta (2b -|r_i - r_j|)$
,where $\Theta(x)$ is the step function,  shows that  for $z= 2,3$ and $z=4$  contacts between monomers that are separated by at least four bonds occur more rapidly as $z$ increases.  The ion-bridging mechanism    
leads  to rapid  condensation of  the chain to metastable 
\textit{compact pearl-necklace   structures}
for  $z=3$ and $4$\cite{kantor,dobrynin}.
 
For $z\geq 2$ the equilibrium globular conformation is kinetically reached provided that the solvent is sufficiently poor.  However, the arrangement of the monomers and counterions in  the globular structure depends  on $v_2$. 
  From Fig.(2a), which shows a plot of the time development of the pair 
distribution function $g(r)$
  for $v_2=-0.34 b^3, l_B = 6b$ and $z=2$,
 we infer that  the combined system of counterions and  the polyelectrolyte forms a Wigner crystal\cite{Wigner}.
The formation of Wigner crystal is seen by the presence of several  peaks
 in $g(r)$ at  long times ($t >350 \tau$). 
The analysis of the peak position reveals that the ordered structure
corresponds to a BCC lattice.  The ordering is considerably fuzzy (see
Fig.(2b)) when the solvent  quality  is extremely poor ($v_2 = -62.34b^3$).   The resulting structure  is 
amorphous (Wigner glass).  Thus, for $v_2 = -62.34b^3$ the dynamics of approach 
to the ordered state is slow i.e., there are manifestations of glassy
dynamics.

\begin{figure}[h]
\leavevmode\centering\psfig{file=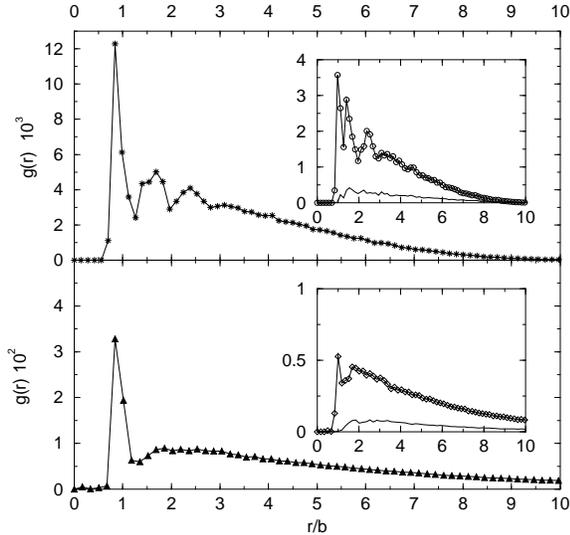,width= 7.5 cm}
\caption{Pair correlation function for the combined system of the PE 
chain ($N=240$) and the counterions ($z=2$) in long time limit ($t >350\tau$).  The system is quenched  to $l_B=6b$.  
The top panel is for $v_2=-0.34b^3$ where the bottom panel 
corresponds to $v_2=-62.34b^3$. The insets show the correlation functions
between monomers (symbols) and between counterions (solid lines). The lack of
correlations between counterions is evident.
}
\label{fig2}
\end{figure}

\begin{figure}[h]
\leavevmode\centering\psfig{file=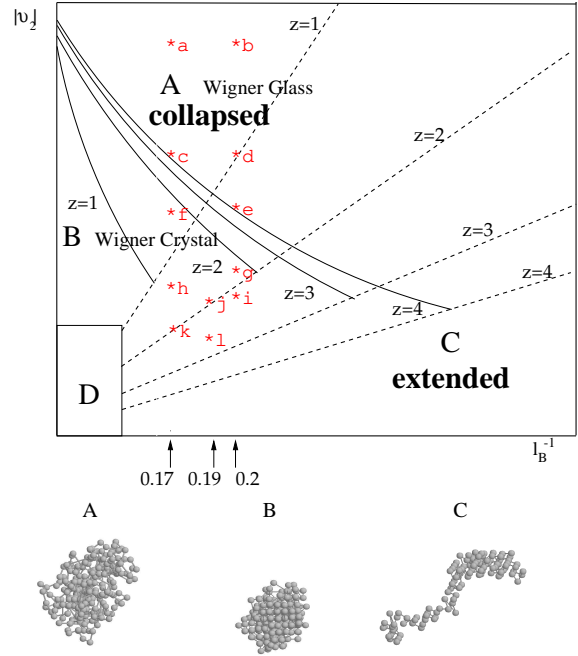,width= 7.5 cm}
\caption{Valence dependent diagram of states in the  ($|v_2|$ and $l_B^{-1}$)
plane for strongly charged PE.  The dashed lines represent the boundary between stretched and collapsed states and depend on $z$.  
The $z$-dependent solid  lines in the collapsed region  separate Wigner crystalline
region from the Wigner ``glassy'' region.  The asteriks are the simulation
results with each data point corresponding to four $z$ values ($z=1-4$). 
Pictures of the conformation of the chain in region A,B and C are also shown.
The $l_B^{-1}$ values are indicated by arrows.  The letter (a-l)  near the 
asterisks corresponds to values of $|v_2|/b^3 =$ 62.34,62.34,25.82,25.82,
15.20,12.36,7.51,4.47,3.91,3.69,0.34,0.06 respectively.  
}
\label{fig3}
\end{figure}

These observations  can be 
used to construct a  valence dependent phase diagram for 
strongly charged polyelectrolyte chain in poor solvents\cite{oosawa}.
Upon counterion condensation  the total charge of the polyelectrolyte decreases from $Nf$  to $N\tilde{f}\approx \kappa (L/l_B)(1/z)$  where $\kappa\approx-ln\phi $  and $\phi$ is the 
volume fraction of the free counterions\cite{schiessel}.  The size of the polyelectrolyte is given by $L\approx \kappa^2b^2N/l_B z^2$ provided 
$l_B> \kappa^2 z^{-2} bN^{1/2}$.
As the quality of the solvent decreases to a level such that 
thermal blob size  $\xi_{T} \approx b^4/|v_2|   < \xi_{el} \approx l_B z^2/\kappa^2$  (size of the electrostatic blob)  then the chain condenses to a globule.
The boundary dividing the stretched and collapsed conformation
is given by $|v_2| \approx b^4\kappa^2/l_B z^2$. 
In the globular  phase we find two regimes, one corresponding to the  Wigner crystal and the other a  Wigner glass.  The boundary between the two is obtained by equating the gain in the  energy upon condensation ($\approx ze^2/d\epsilon$)  to the 
attractive interaction due to the poor solvent quality ($kT v_2^2b^{-6}$). 
This  leads to the condition  $|v_2|\approx  z^{1/2} b^{5/2} l_B^{1/2}$. 
To validate the phase diagram we performed extensive  simulations for all
$z (=1,2,3,4)$ at valuase of $l_B^{-1}$ and $|v_2|$ indicated by asterisks 
in Fig.(\ref{fig3}). The structural conformations are in accord with the predicted phase diagram.

\begin{figure}[h]
\leavevmode\centering\psfig{file=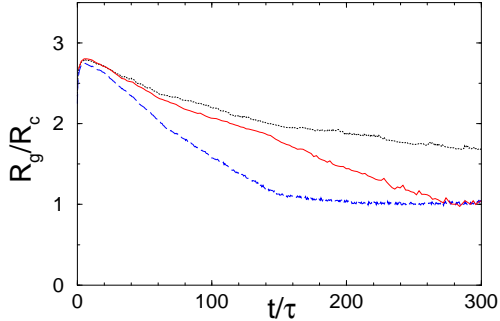,width= 6.5cm}
\caption{The time dependence of $<R_g(t)>  (N=240)$ for different sizes of the counterions ($z=2$) for $v_2=-3.69b^3,l_B=5.3b$.  The top curve (dotted line) is for $ R= 5/7$ and the dashed line is for $R= 2$.  
 For  comparison results for the condition $R=1$ are also displayed as solid line.}
\label{fig4}
\end{figure}

 Oosawa has  argued  that the fraction of condensed counterions is not significantly changed by altering the size of the counterions provided the volume fraction of the polyanion is small\cite{oosawa}.   Since  the correlation between the monomers and counterions  is dependent on the size of the counterions the dynamics of collapse should be sensitive to  $R=\sigma_m/\sigma_c$ where $\sigma_c$ is the size of the counterions.
We carried out simulations for   $ R = 2  $  and $ R = 5/7 $ with $v_2 = -3.69 b^3$   for $N=240$  at $l_B = 5.3b$.
For $R=2$  we find that the rate of globular formation in enhanced compared to   $ R= 1 $  for $z=2, 3$ and  $4$. The difference between  the two cases is most  dramatic for $z=2$ (see Fig.(\ref{fig4})).  
For the divalent cation globular formation  occurs  at  $t\approx 300 \tau$ when $R =1$  (see Fig.(1a)).  However, when the size of counterions is made smaller the PE chain 
almost reaches the globular state  at $ t \approx 150\tau$.  Conversely, when the size of the counterion is increased the efficiency of collapse decreases (see Fig.(\ref{fig4})).   
This establishes the importance of the size of the counterion, and in particular the correlation between them in determining collapse dynamics.

The findings presented here show that a bewildering  range of dynamical 
behavior is to be  expected in  the collapse of strongly charged polyelectrolytes mediated 
by counterion condensation. 
The prediction that the formation of ordered  states  Wigner crystals and amorphous  structure (Wigner glasses) 
should depend on the quality of the solvent should be amenable  to 
experimental (neutron or light scattering) tests.  Even if the Wigner glass
is metastable  our theory  suggests that the dynamics in the two regions
(see Fig.(\ref{fig3})) covering the globular states of PE  should be  dramatically  different and depend 
on the valence of the counterions.  The prediction that size controls 
the kinetics of collapse can  explains  
the changes in the efficiency of DNA condensation with differing cations
 with fixed valence\cite{widom}.

\bf Acknowledgment\rm:
We thank M. Olvera de la Cruz, B.Shklovskii and Y. Levin for discussions.
This work was supported  by a grant from National Science Foundation (NSF CHE 99-75150).


\end{document}